\documentclass[11pt,a4paper,english,superscriptaddress,aps,preprint]{revtex4}
\usepackage{amsfonts}
\usepackage{graphics}
\usepackage{graphicx}
\usepackage{slashed}
\usepackage{amsmath}
\usepackage{amssymb}
\makeatletter
\usepackage{babel}
\newcommand{\bea}{\begin{eqnarray}}
\newcommand{\eea}{\end{eqnarray}}

\newcommand{\pa}{\partial}
\renewcommand{\a}{\alpha}
\newcommand{\ad}{\dot{\alpha}}
\renewcommand{\b}{\beta}
\newcommand{\bd}{\dot{\beta}}
\newcommand{\q}{\theta}

\begin{document}
\immediate\write16{<<WARNING: LINEDRAW macros work with emTeX-dvivers
                    and other drivers supporting emTeX \special's
                    (dviscr, dvihplj, dvidot, dvips, dviwin, etc.) >>}

\newdimen\Lengthunit       \Lengthunit  = 1.5cm
\newcount\Nhalfperiods     \Nhalfperiods= 9
\newcount\magnitude        \magnitude = 1000

\catcode`\*=11
\newdimen\L*   \newdimen\d*   \newdimen\d**
\newdimen\dm*  \newdimen\dd*  \newdimen\dt*
\newdimen\a*   \newdimen\b*   \newdimen\c*
\newdimen\a**  \newdimen\b**
\newdimen\xL*  \newdimen\yL*
\newdimen\rx*  \newdimen\ry*
\newdimen\tmp* \newdimen\linwid*

\newcount\k*   \newcount\l*   \newcount\m*
\newcount\k**  \newcount\l**  \newcount\m**
\newcount\n*   \newcount\dn*  \newcount\r*
\newcount\N*   \newcount\*one \newcount\*two  \*one=1 \*two=2
\newcount\*ths \*ths=1000
\newcount\angle*  \newcount\q*  \newcount\q**
\newcount\angle** \angle**=0
\newcount\sc*     \sc*=0

\newtoks\cos*  \cos*={1}
\newtoks\sin*  \sin*={0}

\catcode`\[=13

\def\rotate(#1){\advance\angle**#1\angle*=\angle**
\q**=\angle*\ifnum\q**<0\q**=-\q**\fi
\ifnum\q**>360\q*=\angle*\divide\q*360\multiply\q*360\advance\angle*-\q*\fi
\ifnum\angle*<0\advance\angle*360\fi\q**=\angle*\divide\q**90\q**=\q**
\def\sgcos*{+}\def\sgsin*{+}\relax
\ifcase\q**\or
 \def\sgcos*{-}\def\sgsin*{+}\or
 \def\sgcos*{-}\def\sgsin*{-}\or
 \def\sgcos*{+}\def\sgsin*{-}\else\fi
\q*=\q**
\multiply\q*90\advance\angle*-\q*
\ifnum\angle*>45\sc*=1\angle*=-\angle*\advance\angle*90\else\sc*=0\fi
\def[##1,##2]{\ifnum\sc*=0\relax
\edef\cs*{\sgcos*.##1}\edef\sn*{\sgsin*.##2}\ifcase\q**\or
 \edef\cs*{\sgcos*.##2}\edef\sn*{\sgsin*.##1}\or
 \edef\cs*{\sgcos*.##1}\edef\sn*{\sgsin*.##2}\or
 \edef\cs*{\sgcos*.##2}\edef\sn*{\sgsin*.##1}\else\fi\else
\edef\cs*{\sgcos*.##2}\edef\sn*{\sgsin*.##1}\ifcase\q**\or
 \edef\cs*{\sgcos*.##1}\edef\sn*{\sgsin*.##2}\or
 \edef\cs*{\sgcos*.##2}\edef\sn*{\sgsin*.##1}\or
 \edef\cs*{\sgcos*.##1}\edef\sn*{\sgsin*.##2}\else\fi\fi
\cos*={\cs*}\sin*={\sn*}\global\edef\gcos*{\cs*}\global\edef\gsin*{\sn*}}\relax
\ifcase\angle*[9999,0]\or
[999,017]\or[999,034]\or[998,052]\or[997,069]\or[996,087]\or
[994,104]\or[992,121]\or[990,139]\or[987,156]\or[984,173]\or
[981,190]\or[978,207]\or[974,224]\or[970,241]\or[965,258]\or
[961,275]\or[956,292]\or[951,309]\or[945,325]\or[939,342]\or
[933,358]\or[927,374]\or[920,390]\or[913,406]\or[906,422]\or
[898,438]\or[891,453]\or[882,469]\or[874,484]\or[866,499]\or
[857,515]\or[848,529]\or[838,544]\or[829,559]\or[819,573]\or
[809,587]\or[798,601]\or[788,615]\or[777,629]\or[766,642]\or
[754,656]\or[743,669]\or[731,681]\or[719,694]\or[707,707]\or
\else[9999,0]\fi}

\catcode`\[=12

\def\GRAPH(hsize=#1)#2{\hbox to #1\Lengthunit{#2\hss}}

\def\Linewidth#1{\global\linwid*=#1\relax
\global\divide\linwid*10\global\multiply\linwid*\mag
\global\divide\linwid*100\special{em:linewidth \the\linwid*}}

\Linewidth{.4pt}
\def\sm*{\special{em:moveto}}
\def\sl*{\special{em:lineto}}
\let\moveto=\sm*
\let\lineto=\sl*
\newbox\spm*   \newbox\spl*
\setbox\spm*\hbox{\sm*}
\setbox\spl*\hbox{\sl*}

\def\mov#1(#2,#3)#4{\rlap{\L*=#1\Lengthunit
\xL*=#2\L* \yL*=#3\L*
\xL*=\xscale\xL* \yL*=\yscale\yL*
\rx* \the\cos*\xL* \tmp* \the\sin*\yL* \advance\rx*-\tmp*
\ry* \the\cos*\yL* \tmp* \the\sin*\xL* \advance\ry*\tmp*
\kern\rx*\raise\ry*\hbox{#4}}}

\def\rmov*(#1,#2)#3{\rlap{\xL*=#1\yL*=#2\relax
\rx* \the\cos*\xL* \tmp* \the\sin*\yL* \advance\rx*-\tmp*
\ry* \the\cos*\yL* \tmp* \the\sin*\xL* \advance\ry*\tmp*
\kern\rx*\raise\ry*\hbox{#3}}}

\def\lin#1(#2,#3){\rlap{\sm*\mov#1(#2,#3){\sl*}}}

\def\arr*(#1,#2,#3){\rmov*(#1\dd*,#1\dt*){\sm*
\rmov*(#2\dd*,#2\dt*){\rmov*(#3\dt*,-#3\dd*){\sl*}}\sm*
\rmov*(#2\dd*,#2\dt*){\rmov*(-#3\dt*,#3\dd*){\sl*}}}}

\def\arrow#1(#2,#3){\rlap{\lin#1(#2,#3)\mov#1(#2,#3){\relax
\d**=-.012\Lengthunit\dd*=#2\d**\dt*=#3\d**
\arr*(1,10,4)\arr*(3,8,4)\arr*(4.8,4.2,3)}}}

\def\arrlin#1(#2,#3){\rlap{\L*=#1\Lengthunit\L*=.5\L*
\lin#1(#2,#3)\rmov*(#2\L*,#3\L*){\arrow.1(#2,#3)}}}

\def\dasharrow#1(#2,#3){\rlap{{\Lengthunit=0.9\Lengthunit
\dashlin#1(#2,#3)\mov#1(#2,#3){\sm*}}\mov#1(#2,#3){\sl*
\d**=-.012\Lengthunit\dd*=#2\d**\dt*=#3\d**
\arr*(1,10,4)\arr*(3,8,4)\arr*(4.8,4.2,3)}}}

\def\clap#1{\hbox to 0pt{\hss #1\hss}}

\def\ind(#1,#2)#3{\rlap{\L*=.1\Lengthunit
\xL*=#1\L* \yL*=#2\L*
\rx* \the\cos*\xL* \tmp* \the\sin*\yL* \advance\rx*-\tmp*
\ry* \the\cos*\yL* \tmp* \the\sin*\xL* \advance\ry*\tmp*
\kern\rx*\raise\ry*\hbox{\lower2pt\clap{$#3$}}}}

\def\sh*(#1,#2)#3{\rlap{\dm*=\the\n*\d**
\xL*=\xscale\dm* \yL*=\yscale\dm* \xL*=#1\xL* \yL*=#2\yL*
\rx* \the\cos*\xL* \tmp* \the\sin*\yL* \advance\rx*-\tmp*
\ry* \the\cos*\yL* \tmp* \the\sin*\xL* \advance\ry*\tmp*
\kern\rx*\raise\ry*\hbox{#3}}}

\def\calcnum*#1(#2,#3){\a*=1000sp\b*=1000sp\a*=#2\a*\b*=#3\b*
\ifdim\a*<0pt\a*-\a*\fi\ifdim\b*<0pt\b*-\b*\fi
\ifdim\a*>\b*\c*=.96\a*\advance\c*.4\b*
\else\c*=.96\b*\advance\c*.4\a*\fi
\k*\a*\multiply\k*\k*\l*\b*\multiply\l*\l*
\m*\k*\advance\m*\l*\n*\c*\r*\n*\multiply\n*\n*
\dn*\m*\advance\dn*-\n*\divide\dn*2\divide\dn*\r*
\advance\r*\dn*
\c*=\the\Nhalfperiods5sp\c*=#1\c*\ifdim\c*<0pt\c*-\c*\fi
\multiply\c*\r*\N*\c*\divide\N*10000}

\def\dashlin#1(#2,#3){\rlap{\calcnum*#1(#2,#3)\relax
\d**=#1\Lengthunit\ifdim\d**<0pt\d**-\d**\fi
\divide\N*2\multiply\N*2\advance\N*\*one
\divide\d**\N*\sm*\n*\*one\sh*(#2,#3){\sl*}\loop
\advance\n*\*one\sh*(#2,#3){\sm*}\advance\n*\*one
\sh*(#2,#3){\sl*}\ifnum\n*<\N*\repeat}}

\def\dashdotlin#1(#2,#3){\rlap{\calcnum*#1(#2,#3)\relax
\d**=#1\Lengthunit\ifdim\d**<0pt\d**-\d**\fi
\divide\N*2\multiply\N*2\advance\N*1\multiply\N*2\relax
\divide\d**\N*\sm*\n*\*two\sh*(#2,#3){\sl*}\loop
\advance\n*\*one\sh*(#2,#3){\kern-1.48pt\lower.5pt\hbox{\rm.}}\relax
\advance\n*\*one\sh*(#2,#3){\sm*}\advance\n*\*two
\sh*(#2,#3){\sl*}\ifnum\n*<\N*\repeat}}

\def\shl*(#1,#2)#3{\kern#1#3\lower#2#3\hbox{\unhcopy\spl*}}

\def\trianglin#1(#2,#3){\rlap{\toks0={#2}\toks1={#3}\calcnum*#1(#2,#3)\relax
\dd*=.57\Lengthunit\dd*=#1\dd*\divide\dd*\N*
\divide\dd*\*ths \multiply\dd*\magnitude
\d**=#1\Lengthunit\ifdim\d**<0pt\d**-\d**\fi
\multiply\N*2\divide\d**\N*\sm*\n*\*one\loop
\shl**{\dd*}\dd*-\dd*\advance\n*2\relax
\ifnum\n*<\N*\repeat\n*\N*\shl**{0pt}}}

\def\wavelin#1(#2,#3){\rlap{\toks0={#2}\toks1={#3}\calcnum*#1(#2,#3)\relax
\dd*=.23\Lengthunit\dd*=#1\dd*\divide\dd*\N*
\divide\dd*\*ths \multiply\dd*\magnitude
\d**=#1\Lengthunit\ifdim\d**<0pt\d**-\d**\fi
\multiply\N*4\divide\d**\N*\sm*\n*\*one\loop
\shl**{\dd*}\dt*=1.3\dd*\advance\n*\*one
\shl**{\dt*}\advance\n*\*one
\shl**{\dd*}\advance\n*\*two
\dd*-\dd*\ifnum\n*<\N*\repeat\n*\N*\shl**{0pt}}}

\def\w*lin(#1,#2){\rlap{\toks0={#1}\toks1={#2}\d**=\Lengthunit\dd*=-.12\d**
\divide\dd*\*ths \multiply\dd*\magnitude
\N*8\divide\d**\N*\sm*\n*\*one\loop
\shl**{\dd*}\dt*=1.3\dd*\advance\n*\*one
\shl**{\dt*}\advance\n*\*one
\shl**{\dd*}\advance\n*\*one
\shl**{0pt}\dd*-\dd*\advance\n*1\ifnum\n*<\N*\repeat}}

\def\l*arc(#1,#2)[#3][#4]{\rlap{\toks0={#1}\toks1={#2}\d**=\Lengthunit
\dd*=#3.037\d**\dd*=#4\dd*\dt*=#3.049\d**\dt*=#4\dt*\ifdim\d**>10mm\relax
\d**=.25\d**\n*\*one\shl**{-\dd*}\n*\*two\shl**{-\dt*}\n*3\relax
\shl**{-\dd*}\n*4\relax\shl**{0pt}\else
\ifdim\d**>5mm\d**=.5\d**\n*\*one\shl**{-\dt*}\n*\*two
\shl**{0pt}\else\n*\*one\shl**{0pt}\fi\fi}}

\def\d*arc(#1,#2)[#3][#4]{\rlap{\toks0={#1}\toks1={#2}\d**=\Lengthunit
\dd*=#3.037\d**\dd*=#4\dd*\d**=.25\d**\sm*\n*\*one\shl**{-\dd*}\relax
\n*3\relax\sh*(#1,#2){\xL*=\xscale\dd*\yL*=\yscale\dd*
\kern#2\xL*\lower#1\yL*\hbox{\sm*}}\n*4\relax\shl**{0pt}}}

\def\shl**#1{\c*=\the\n*\d**\d*=#1\relax
\a*=\the\toks0\c*\b*=\the\toks1\d*\advance\a*-\b*
\b*=\the\toks1\c*\d*=\the\toks0\d*\advance\b*\d*
\a*=\xscale\a*\b*=\yscale\b*
\rx* \the\cos*\a* \tmp* \the\sin*\b* \advance\rx*-\tmp*
\ry* \the\cos*\b* \tmp* \the\sin*\a* \advance\ry*\tmp*
\raise\ry*\rlap{\kern\rx*\unhcopy\spl*}}

\def\wlin*#1(#2,#3)[#4]{\rlap{\toks0={#2}\toks1={#3}\relax
\c*=#1\l*\c*\c*=.01\Lengthunit\m*\c*\divide\l*\m*
\c*=\the\Nhalfperiods5sp\multiply\c*\l*\N*\c*\divide\N*\*ths
\divide\N*2\multiply\N*2\advance\N*\*one
\dd*=.002\Lengthunit\dd*=#4\dd*\multiply\dd*\l*\divide\dd*\N*
\divide\dd*\*ths \multiply\dd*\magnitude
\d**=#1\multiply\N*4\divide\d**\N*\sm*\n*\*one\loop
\shl**{\dd*}\dt*=1.3\dd*\advance\n*\*one
\shl**{\dt*}\advance\n*\*one
\shl**{\dd*}\advance\n*\*two
\dd*-\dd*\ifnum\n*<\N*\repeat\n*\N*\shl**{0pt}}}

\def\wavebox#1{\setbox0\hbox{#1}\relax
\a*=\wd0\advance\a*14pt\b*=\ht0\advance\b*\dp0\advance\b*14pt\relax
\hbox{\kern9pt\relax
\rmov*(0pt,\ht0){\rmov*(-7pt,7pt){\wlin*\a*(1,0)[+]\wlin*\b*(0,-1)[-]}}\relax
\rmov*(\wd0,-\dp0){\rmov*(7pt,-7pt){\wlin*\a*(-1,0)[+]\wlin*\b*(0,1)[-]}}\relax
\box0\kern9pt}}

\def\rectangle#1(#2,#3){\relax
\lin#1(#2,0)\lin#1(0,#3)\mov#1(0,#3){\lin#1(#2,0)}\mov#1(#2,0){\lin#1(0,#3)}}

\def\dashrectangle#1(#2,#3){\dashlin#1(#2,0)\dashlin#1(0,#3)\relax
\mov#1(0,#3){\dashlin#1(#2,0)}\mov#1(#2,0){\dashlin#1(0,#3)}}

\def\waverectangle#1(#2,#3){\L*=#1\Lengthunit\a*=#2\L*\b*=#3\L*
\ifdim\a*<0pt\a*-\a*\def\x*{-1}\else\def\x*{1}\fi
\ifdim\b*<0pt\b*-\b*\def\y*{-1}\else\def\y*{1}\fi
\wlin*\a*(\x*,0)[-]\wlin*\b*(0,\y*)[+]\relax
\mov#1(0,#3){\wlin*\a*(\x*,0)[+]}\mov#1(#2,0){\wlin*\b*(0,\y*)[-]}}

\def\calcparab*{\ifnum\n*>\m*\k*\N*\advance\k*-\n*\else\k*\n*\fi
\a*=\the\k* sp\a*=10\a*\b*\dm*\advance\b*-\a*\k*\b*
\a*=\the\*ths\b*\divide\a*\l*\multiply\a*\k*
\divide\a*\l*\k*\*ths\r*\a*\advance\k*-\r*\dt*=\the\k*\L*}

\def\arcto#1(#2,#3)[#4]{\rlap{\toks0={#2}\toks1={#3}\calcnum*#1(#2,#3)\relax
\dm*=135sp\dm*=#1\dm*\d**=#1\Lengthunit\ifdim\dm*<0pt\dm*-\dm*\fi
\multiply\dm*\r*\a*=.3\dm*\a*=#4\a*\ifdim\a*<0pt\a*-\a*\fi
\advance\dm*\a*\N*\dm*\divide\N*10000\relax
\divide\N*2\multiply\N*2\advance\N*\*one
\L*=-.25\d**\L*=#4\L*\divide\d**\N*\divide\L*\*ths
\m*\N*\divide\m*2\dm*=\the\m*5sp\l*\dm*\sm*\n*\*one\loop
\calcparab*\shl**{-\dt*}\advance\n*1\ifnum\n*<\N*\repeat}}

\def\arrarcto#1(#2,#3)[#4]{\L*=#1\Lengthunit\L*=.54\L*
\arcto#1(#2,#3)[#4]\rmov*(#2\L*,#3\L*){\d*=.457\L*\d*=#4\d*\d**-\d*
\rmov*(#3\d**,#2\d*){\arrow.02(#2,#3)}}}

\def\dasharcto#1(#2,#3)[#4]{\rlap{\toks0={#2}\toks1={#3}\relax
\calcnum*#1(#2,#3)\dm*=\the\N*5sp\a*=.3\dm*\a*=#4\a*\ifdim\a*<0pt\a*-\a*\fi
\advance\dm*\a*\N*\dm*
\divide\N*20\multiply\N*2\advance\N*1\d**=#1\Lengthunit
\L*=-.25\d**\L*=#4\L*\divide\d**\N*\divide\L*\*ths
\m*\N*\divide\m*2\dm*=\the\m*5sp\l*\dm*
\sm*\n*\*one\loop\calcparab*
\shl**{-\dt*}\advance\n*1\ifnum\n*>\N*\else\calcparab*
\sh*(#2,#3){\xL*=#3\dt* \yL*=#2\dt*
\rx* \the\cos*\xL* \tmp* \the\sin*\yL* \advance\rx*\tmp*
\ry* \the\cos*\yL* \tmp* \the\sin*\xL* \advance\ry*-\tmp*
\kern\rx*\lower\ry*\hbox{\sm*}}\fi
\advance\n*1\ifnum\n*<\N*\repeat}}

\def\*shl*#1{\c*=\the\n*\d**\advance\c*#1\a**\d*\dt*\advance\d*#1\b**
\a*=\the\toks0\c*\b*=\the\toks1\d*\advance\a*-\b*
\b*=\the\toks1\c*\d*=\the\toks0\d*\advance\b*\d*
\rx* \the\cos*\a* \tmp* \the\sin*\b* \advance\rx*-\tmp*
\ry* \the\cos*\b* \tmp* \the\sin*\a* \advance\ry*\tmp*
\raise\ry*\rlap{\kern\rx*\unhcopy\spl*}}

\def\calcnormal*#1{\b**=10000sp\a**\b**\k*\n*\advance\k*-\m*
\multiply\a**\k*\divide\a**\m*\a**=#1\a**\ifdim\a**<0pt\a**-\a**\fi
\ifdim\a**>\b**\d*=.96\a**\advance\d*.4\b**
\else\d*=.96\b**\advance\d*.4\a**\fi
\d*=.01\d*\r*\d*\divide\a**\r*\divide\b**\r*
\ifnum\k*<0\a**-\a**\fi\d*=#1\d*\ifdim\d*<0pt\b**-\b**\fi
\k*\a**\a**=\the\k*\dd*\k*\b**\b**=\the\k*\dd*}

\def\wavearcto#1(#2,#3)[#4]{\rlap{\toks0={#2}\toks1={#3}\relax
\calcnum*#1(#2,#3)\c*=\the\N*5sp\a*=.4\c*\a*=#4\a*\ifdim\a*<0pt\a*-\a*\fi
\advance\c*\a*\N*\c*\divide\N*20\multiply\N*2\advance\N*-1\multiply\N*4\relax
\d**=#1\Lengthunit\dd*=.012\d**
\divide\dd*\*ths \multiply\dd*\magnitude
\ifdim\d**<0pt\d**-\d**\fi\L*=.25\d**
\divide\d**\N*\divide\dd*\N*\L*=#4\L*\divide\L*\*ths
\m*\N*\divide\m*2\dm*=\the\m*0sp\l*\dm*
\sm*\n*\*one\loop\calcnormal*{#4}\calcparab*
\*shl*{1}\advance\n*\*one\calcparab*
\*shl*{1.3}\advance\n*\*one\calcparab*
\*shl*{1}\advance\n*2\dd*-\dd*\ifnum\n*<\N*\repeat\n*\N*\shl**{0pt}}}

\def\triangarcto#1(#2,#3)[#4]{\rlap{\toks0={#2}\toks1={#3}\relax
\calcnum*#1(#2,#3)\c*=\the\N*5sp\a*=.4\c*\a*=#4\a*\ifdim\a*<0pt\a*-\a*\fi
\advance\c*\a*\N*\c*\divide\N*20\multiply\N*2\advance\N*-1\multiply\N*2\relax
\d**=#1\Lengthunit\dd*=.012\d**
\divide\dd*\*ths \multiply\dd*\magnitude
\ifdim\d**<0pt\d**-\d**\fi\L*=.25\d**
\divide\d**\N*\divide\dd*\N*\L*=#4\L*\divide\L*\*ths
\m*\N*\divide\m*2\dm*=\the\m*0sp\l*\dm*
\sm*\n*\*one\loop\calcnormal*{#4}\calcparab*
\*shl*{1}\advance\n*2\dd*-\dd*\ifnum\n*<\N*\repeat\n*\N*\shl**{0pt}}}

\def\hr*#1{\L*=\xscale\Lengthunit\ifnum
\angle**=0\clap{\vrule width#1\L* height.1pt}\else
\L*=#1\L*\L*=.5\L*\rmov*(-\L*,0pt){\sm*}\rmov*(\L*,0pt){\sl*}\fi}

\def\shade#1[#2]{\rlap{\Lengthunit=#1\Lengthunit
\special{em:linewidth .001pt}\relax
\mov(0,#2.05){\hr*{.994}}\mov(0,#2.1){\hr*{.980}}\relax
\mov(0,#2.15){\hr*{.953}}\mov(0,#2.2){\hr*{.916}}\relax
\mov(0,#2.25){\hr*{.867}}\mov(0,#2.3){\hr*{.798}}\relax
\mov(0,#2.35){\hr*{.715}}\mov(0,#2.4){\hr*{.603}}\relax
\mov(0,#2.45){\hr*{.435}}\special{em:linewidth \the\linwid*}}}

\def\dshade#1[#2]{\rlap{\special{em:linewidth .001pt}\relax
\Lengthunit=#1\Lengthunit\if#2-\def\t*{+}\else\def\t*{-}\fi
\mov(0,\t*.025){\relax
\mov(0,#2.05){\hr*{.995}}\mov(0,#2.1){\hr*{.988}}\relax
\mov(0,#2.15){\hr*{.969}}\mov(0,#2.2){\hr*{.937}}\relax
\mov(0,#2.25){\hr*{.893}}\mov(0,#2.3){\hr*{.836}}\relax
\mov(0,#2.35){\hr*{.760}}\mov(0,#2.4){\hr*{.662}}\relax
\mov(0,#2.45){\hr*{.531}}\mov(0,#2.5){\hr*{.320}}\relax
\special{em:linewidth \the\linwid*}}}}

\def\vdot{\rlap{\kern-1.9pt\lower1.8pt\hbox{$\scriptstyle\bullet$}}}
\def\vtimes{\rlap{\kern-3pt\lower1.8pt\hbox{$\scriptstyle\times$}}}
\def\vDot{\rlap{\kern-2.3pt\lower2.7pt\hbox{$\bullet$}}}
\def\vTimes{\rlap{\kern-3.6pt\lower2.4pt\hbox{$\times$}}}

\def\arc(#1)[#2,#3]{{\k*=#2\l*=#3\m*=\l*
\advance\m*-6\ifnum\k*>\l*\relax\else
{\rotate(#2)\mov(#1,0){\sm*}}\loop
\ifnum\k*<\m*\advance\k*5{\rotate(\k*)\mov(#1,0){\sl*}}\repeat
{\rotate(#3)\mov(#1,0){\sl*}}\fi}}

\def\dasharc(#1)[#2,#3]{{\k**=#2\n*=#3\advance\n*-1\advance\n*-\k**
\L*=1000sp\L*#1\L* \multiply\L*\n* \multiply\L*\Nhalfperiods
\divide\L*57\N*\L* \divide\N*2000\ifnum\N*=0\N*1\fi
\r*\n*  \divide\r*\N* \ifnum\r*<2\r*2\fi
\m**\r* \divide\m**2 \l**\r* \advance\l**-\m** \N*\n* \divide\N*\r*
\k**\r* \multiply\k**\N* \dn*\n*
\advance\dn*-\k** \divide\dn*2\advance\dn*\*one
\r*\l** \divide\r*2\advance\dn*\r* \advance\N*-2\k**#2\relax
\ifnum\l**<6{\rotate(#2)\mov(#1,0){\sm*}}\advance\k**\dn*
{\rotate(\k**)\mov(#1,0){\sl*}}\advance\k**\m**
{\rotate(\k**)\mov(#1,0){\sm*}}\loop
\advance\k**\l**{\rotate(\k**)\mov(#1,0){\sl*}}\advance\k**\m**
{\rotate(\k**)\mov(#1,0){\sm*}}\advance\N*-1\ifnum\N*>0\repeat
{\rotate(#3)\mov(#1,0){\sl*}}\else\advance\k**\dn*
\arc(#1)[#2,\k**]\loop\advance\k**\m** \r*\k**
\advance\k**\l** {\arc(#1)[\r*,\k**]}\relax
\advance\N*-1\ifnum\N*>0\repeat
\advance\k**\m**\arc(#1)[\k**,#3]\fi}}

\def\triangarc#1(#2)[#3,#4]{{\k**=#3\n*=#4\advance\n*-\k**
\L*=1000sp\L*#2\L* \multiply\L*\n* \multiply\L*\Nhalfperiods
\divide\L*57\N*\L* \divide\N*1000\ifnum\N*=0\N*1\fi
\d**=#2\Lengthunit \d*\d** \divide\d*57\multiply\d*\n*
\r*\n*  \divide\r*\N* \ifnum\r*<2\r*2\fi
\m**\r* \divide\m**2 \l**\r* \advance\l**-\m** \N*\n* \divide\N*\r*
\dt*\d* \divide\dt*\N* \dt*.5\dt* \dt*#1\dt*
\divide\dt*1000\multiply\dt*\magnitude
\k**\r* \multiply\k**\N* \dn*\n* \advance\dn*-\k** \divide\dn*2\relax
\r*\l** \divide\r*2\advance\dn*\r* \advance\N*-1\k**#3\relax
{\rotate(#3)\mov(#2,0){\sm*}}\advance\k**\dn*
{\rotate(\k**)\mov(#2,0){\sl*}}\advance\k**-\m**\advance\l**\m**\loop\dt*-\dt*
\d*\d** \advance\d*\dt*
\advance\k**\l**{\rotate(\k**)\rmov*(\d*,0pt){\sl*}}%
\advance\N*-1\ifnum\N*>0\repeat\advance\k**\m**
{\rotate(\k**)\mov(#2,0){\sl*}}{\rotate(#4)\mov(#2,0){\sl*}}}}

\def\wavearc#1(#2)[#3,#4]{{\k**=#3\n*=#4\advance\n*-\k**
\L*=4000sp\L*#2\L* \multiply\L*\n* \multiply\L*\Nhalfperiods
\divide\L*57\N*\L* \divide\N*1000\ifnum\N*=0\N*1\fi
\d**=#2\Lengthunit \d*\d** \divide\d*57\multiply\d*\n*
\r*\n*  \divide\r*\N* \ifnum\r*=0\r*1\fi
\m**\r* \divide\m**2 \l**\r* \advance\l**-\m** \N*\n* \divide\N*\r*
\dt*\d* \divide\dt*\N* \dt*.7\dt* \dt*#1\dt*
\divide\dt*1000\multiply\dt*\magnitude
\k**\r* \multiply\k**\N* \dn*\n* \advance\dn*-\k** \divide\dn*2\relax
\divide\N*4\advance\N*-1\k**#3\relax
{\rotate(#3)\mov(#2,0){\sm*}}\advance\k**\dn*
{\rotate(\k**)\mov(#2,0){\sl*}}\advance\k**-\m**\advance\l**\m**\loop\dt*-\dt*
\d*\d** \advance\d*\dt* \dd*\d** \advance\dd*1.3\dt*
\advance\k**\r*{\rotate(\k**)\rmov*(\d*,0pt){\sl*}}\relax
\advance\k**\r*{\rotate(\k**)\rmov*(\dd*,0pt){\sl*}}\relax
\advance\k**\r*{\rotate(\k**)\rmov*(\d*,0pt){\sl*}}\relax
\advance\k**\r*
\advance\N*-1\ifnum\N*>0\repeat\advance\k**\m**
{\rotate(\k**)\mov(#2,0){\sl*}}{\rotate(#4)\mov(#2,0){\sl*}}}}

\def\gmov*#1(#2,#3)#4{\rlap{\L*=#1\Lengthunit
\xL*=#2\L* \yL*=#3\L*
\rx* \gcos*\xL* \tmp* \gsin*\yL* \advance\rx*-\tmp*
\ry* \gcos*\yL* \tmp* \gsin*\xL* \advance\ry*\tmp*
\rx*=\xscale\rx* \ry*=\yscale\ry*
\xL* \the\cos*\rx* \tmp* \the\sin*\ry* \advance\xL*-\tmp*
\yL* \the\cos*\ry* \tmp* \the\sin*\rx* \advance\yL*\tmp*
\kern\xL*\raise\yL*\hbox{#4}}}

\def\rgmov*(#1,#2)#3{\rlap{\xL*#1\yL*#2\relax
\rx* \gcos*\xL* \tmp* \gsin*\yL* \advance\rx*-\tmp*
\ry* \gcos*\yL* \tmp* \gsin*\xL* \advance\ry*\tmp*
\rx*=\xscale\rx* \ry*=\yscale\ry*
\xL* \the\cos*\rx* \tmp* \the\sin*\ry* \advance\xL*-\tmp*
\yL* \the\cos*\ry* \tmp* \the\sin*\rx* \advance\yL*\tmp*
\kern\xL*\raise\yL*\hbox{#3}}}

\def\Earc(#1)[#2,#3][#4,#5]{{\k*=#2\l*=#3\m*=\l*
\advance\m*-6\ifnum\k*>\l*\relax\else\def\xscale{#4}\def\yscale{#5}\relax
{\angle**0\rotate(#2)}\gmov*(#1,0){\sm*}\loop
\ifnum\k*<\m*\advance\k*5\relax
{\angle**0\rotate(\k*)}\gmov*(#1,0){\sl*}\repeat
{\angle**0\rotate(#3)}\gmov*(#1,0){\sl*}\relax
\def\xscale{1}\def\yscale{1}\fi}}

\def\dashEarc(#1)[#2,#3][#4,#5]{{\k**=#2\n*=#3\advance\n*-1\advance\n*-\k**
\L*=1000sp\L*#1\L* \multiply\L*\n* \multiply\L*\Nhalfperiods
\divide\L*57\N*\L* \divide\N*2000\ifnum\N*=0\N*1\fi
\r*\n*  \divide\r*\N* \ifnum\r*<2\r*2\fi
\m**\r* \divide\m**2 \l**\r* \advance\l**-\m** \N*\n* \divide\N*\r*
\k**\r*\multiply\k**\N* \dn*\n* \advance\dn*-\k** \divide\dn*2\advance\dn*\*one
\r*\l** \divide\r*2\advance\dn*\r* \advance\N*-2\k**#2\relax
\ifnum\l**<6\def\xscale{#4}\def\yscale{#5}\relax
{\angle**0\rotate(#2)}\gmov*(#1,0){\sm*}\advance\k**\dn*
{\angle**0\rotate(\k**)}\gmov*(#1,0){\sl*}\advance\k**\m**
{\angle**0\rotate(\k**)}\gmov*(#1,0){\sm*}\loop
\advance\k**\l**{\angle**0\rotate(\k**)}\gmov*(#1,0){\sl*}\advance\k**\m**
{\angle**0\rotate(\k**)}\gmov*(#1,0){\sm*}\advance\N*-1\ifnum\N*>0\repeat
{\angle**0\rotate(#3)}\gmov*(#1,0){\sl*}\def\xscale{1}\def\yscale{1}\else
\advance\k**\dn* \Earc(#1)[#2,\k**][#4,#5]\loop\advance\k**\m** \r*\k**
\advance\k**\l** {\Earc(#1)[\r*,\k**][#4,#5]}\relax
\advance\N*-1\ifnum\N*>0\repeat
\advance\k**\m**\Earc(#1)[\k**,#3][#4,#5]\fi}}

\def\triangEarc#1(#2)[#3,#4][#5,#6]{{\k**=#3\n*=#4\advance\n*-\k**
\L*=1000sp\L*#2\L* \multiply\L*\n* \multiply\L*\Nhalfperiods
\divide\L*57\N*\L* \divide\N*1000\ifnum\N*=0\N*1\fi
\d**=#2\Lengthunit \d*\d** \divide\d*57\multiply\d*\n*
\r*\n*  \divide\r*\N* \ifnum\r*<2\r*2\fi
\m**\r* \divide\m**2 \l**\r* \advance\l**-\m** \N*\n* \divide\N*\r*
\dt*\d* \divide\dt*\N* \dt*.5\dt* \dt*#1\dt*
\divide\dt*1000\multiply\dt*\magnitude
\k**\r* \multiply\k**\N* \dn*\n* \advance\dn*-\k** \divide\dn*2\relax
\r*\l** \divide\r*2\advance\dn*\r* \advance\N*-1\k**#3\relax
\def\xscale{#5}\def\yscale{#6}\relax
{\angle**0\rotate(#3)}\gmov*(#2,0){\sm*}\advance\k**\dn*
{\angle**0\rotate(\k**)}\gmov*(#2,0){\sl*}\advance\k**-\m**
\advance\l**\m**\loop\dt*-\dt* \d*\d** \advance\d*\dt*
\advance\k**\l**{\angle**0\rotate(\k**)}\rgmov*(\d*,0pt){\sl*}\relax
\advance\N*-1\ifnum\N*>0\repeat\advance\k**\m**
{\angle**0\rotate(\k**)}\gmov*(#2,0){\sl*}\relax
{\angle**0\rotate(#4)}\gmov*(#2,0){\sl*}\def\xscale{1}\def\yscale{1}}}

\def\waveEarc#1(#2)[#3,#4][#5,#6]{{\k**=#3\n*=#4\advance\n*-\k**
\L*=4000sp\L*#2\L* \multiply\L*\n* \multiply\L*\Nhalfperiods
\divide\L*57\N*\L* \divide\N*1000\ifnum\N*=0\N*1\fi
\d**=#2\Lengthunit \d*\d** \divide\d*57\multiply\d*\n*
\r*\n*  \divide\r*\N* \ifnum\r*=0\r*1\fi
\m**\r* \divide\m**2 \l**\r* \advance\l**-\m** \N*\n* \divide\N*\r*
\dt*\d* \divide\dt*\N* \dt*.7\dt* \dt*#1\dt*
\divide\dt*1000\multiply\dt*\magnitude
\k**\r* \multiply\k**\N* \dn*\n* \advance\dn*-\k** \divide\dn*2\relax
\divide\N*4\advance\N*-1\k**#3\def\xscale{#5}\def\yscale{#6}\relax
{\angle**0\rotate(#3)}\gmov*(#2,0){\sm*}\advance\k**\dn*
{\angle**0\rotate(\k**)}\gmov*(#2,0){\sl*}\advance\k**-\m**
\advance\l**\m**\loop\dt*-\dt*
\d*\d** \advance\d*\dt* \dd*\d** \advance\dd*1.3\dt*
\advance\k**\r*{\angle**0\rotate(\k**)}\rgmov*(\d*,0pt){\sl*}\relax
\advance\k**\r*{\angle**0\rotate(\k**)}\rgmov*(\dd*,0pt){\sl*}\relax
\advance\k**\r*{\angle**0\rotate(\k**)}\rgmov*(\d*,0pt){\sl*}\relax
\advance\k**\r*
\advance\N*-1\ifnum\N*>0\repeat\advance\k**\m**
{\angle**0\rotate(\k**)}\gmov*(#2,0){\sl*}\relax
{\angle**0\rotate(#4)}\gmov*(#2,0){\sl*}\def\xscale{1}\def\yscale{1}}}

\newcount\CatcodeOfAtSign
\CatcodeOfAtSign=\the\catcode`\@
\catcode`\@=11
\def\@arc#1[#2][#3]{\rlap{\Lengthunit=#1\Lengthunit
\sm*\l*arc(#2.1914,#3.0381)[#2][#3]\relax
\mov(#2.1914,#3.0381){\l*arc(#2.1622,#3.1084)[#2][#3]}\relax
\mov(#2.3536,#3.1465){\l*arc(#2.1084,#3.1622)[#2][#3]}\relax
\mov(#2.4619,#3.3086){\l*arc(#2.0381,#3.1914)[#2][#3]}}}

\def\dash@arc#1[#2][#3]{\rlap{\Lengthunit=#1\Lengthunit
\d*arc(#2.1914,#3.0381)[#2][#3]\relax
\mov(#2.1914,#3.0381){\d*arc(#2.1622,#3.1084)[#2][#3]}\relax
\mov(#2.3536,#3.1465){\d*arc(#2.1084,#3.1622)[#2][#3]}\relax
\mov(#2.4619,#3.3086){\d*arc(#2.0381,#3.1914)[#2][#3]}}}

\def\wave@arc#1[#2][#3]{\rlap{\Lengthunit=#1\Lengthunit
\w*lin(#2.1914,#3.0381)\relax
\mov(#2.1914,#3.0381){\w*lin(#2.1622,#3.1084)}\relax
\mov(#2.3536,#3.1465){\w*lin(#2.1084,#3.1622)}\relax
\mov(#2.4619,#3.3086){\w*lin(#2.0381,#3.1914)}}}

\def\bezier#1(#2,#3)(#4,#5)(#6,#7){\N*#1\l*\N* \advance\l*\*one
\d* #4\Lengthunit \advance\d* -#2\Lengthunit \multiply\d* \*two
\b* #6\Lengthunit \advance\b* -#2\Lengthunit
\advance\b*-\d* \divide\b*\N*
\d** #5\Lengthunit \advance\d** -#3\Lengthunit \multiply\d** \*two
\b** #7\Lengthunit \advance\b** -#3\Lengthunit
\advance\b** -\d** \divide\b**\N*
\mov(#2,#3){\sm*{\loop\ifnum\m*<\l*
\a*\m*\b* \advance\a*\d* \divide\a*\N* \multiply\a*\m*
\a**\m*\b** \advance\a**\d** \divide\a**\N* \multiply\a**\m*
\rmov*(\a*,\a**){\unhcopy\spl*}\advance\m*\*one\repeat}}}

\catcode`\*=12

\newcount\n@ast

\def\n@ast@#1{\n@ast0\relax\get@ast@#1\end}
\def\get@ast@#1{\ifx#1\end\let\next\relax\else
\ifx#1*\advance\n@ast1\fi\let\next\get@ast@\fi\next}

\newif\if@up \newif\if@dwn
\def\up@down@#1{\@upfalse\@dwnfalse
\if#1u\@uptrue\fi\if#1U\@uptrue\fi\if#1+\@uptrue\fi
\if#1d\@dwntrue\fi\if#1D\@dwntrue\fi\if#1-\@dwntrue\fi}

\def\halfcirc#1(#2)[#3]{{\Lengthunit=#2\Lengthunit\up@down@{#3}\relax
\if@up\mov(0,.5){\@arc[-][-]\@arc[+][-]}\fi
\if@dwn\mov(0,-.5){\@arc[-][+]\@arc[+][+]}\fi
\def\lft{\mov(0,.5){\@arc[-][-]}\mov(0,-.5){\@arc[-][+]}}\relax
\def\rght{\mov(0,.5){\@arc[+][-]}\mov(0,-.5){\@arc[+][+]}}\relax
\if#3l\lft\fi\if#3L\lft\fi\if#3r\rght\fi\if#3R\rght\fi
\n@ast@{#1}\relax
\ifnum\n@ast>0\if@up\shade[+]\fi\if@dwn\shade[-]\fi\fi
\ifnum\n@ast>1\if@up\dshade[+]\fi\if@dwn\dshade[-]\fi\fi}}

\def\halfdashcirc(#1)[#2]{{\Lengthunit=#1\Lengthunit\up@down@{#2}\relax
\if@up\mov(0,.5){\dash@arc[-][-]\dash@arc[+][-]}\fi
\if@dwn\mov(0,-.5){\dash@arc[-][+]\dash@arc[+][+]}\fi
\def\lft{\mov(0,.5){\dash@arc[-][-]}\mov(0,-.5){\dash@arc[-][+]}}\relax
\def\rght{\mov(0,.5){\dash@arc[+][-]}\mov(0,-.5){\dash@arc[+][+]}}\relax
\if#2l\lft\fi\if#2L\lft\fi\if#2r\rght\fi\if#2R\rght\fi}}

\def\halfwavecirc(#1)[#2]{{\Lengthunit=#1\Lengthunit\up@down@{#2}\relax
\if@up\mov(0,.5){\wave@arc[-][-]\wave@arc[+][-]}\fi
\if@dwn\mov(0,-.5){\wave@arc[-][+]\wave@arc[+][+]}\fi
\def\lft{\mov(0,.5){\wave@arc[-][-]}\mov(0,-.5){\wave@arc[-][+]}}\relax
\def\rght{\mov(0,.5){\wave@arc[+][-]}\mov(0,-.5){\wave@arc[+][+]}}\relax
\if#2l\lft\fi\if#2L\lft\fi\if#2r\rght\fi\if#2R\rght\fi}}

\catcode`\*=11

\def\Circle#1(#2){\halfcirc#1(#2)[u]\halfcirc#1(#2)[d]\n@ast@{#1}\relax
\ifnum\n@ast>0\L*=\xscale\Lengthunit
\ifnum\angle**=0\clap{\vrule width#2\L* height.1pt}\else
\L*=#2\L*\L*=.5\L*\special{em:linewidth .001pt}\relax
\rmov*(-\L*,0pt){\sm*}\rmov*(\L*,0pt){\sl*}\relax
\special{em:linewidth \the\linwid*}\fi\fi}

\catcode`\*=12

\def\wavecirc(#1){\halfwavecirc(#1)[u]\halfwavecirc(#1)[d]}
\def\dashcirc(#1){\halfdashcirc(#1)[u]\halfdashcirc(#1)[d]}

\def\xscale{1}

\def\yscale{1}

\def\Ellipse#1(#2)[#3,#4]{\def\xscale{#3}\def\yscale{#4}\relax
\Circle#1(#2)\def\xscale{1}\def\yscale{1}}

\def\dashEllipse(#1)[#2,#3]{\def\xscale{#2}\def\yscale{#3}\relax
\dashcirc(#1)\def\xscale{1}\def\yscale{1}}

\def\waveEllipse(#1)[#2,#3]{\def\xscale{#2}\def\yscale{#3}\relax
\wavecirc(#1)\def\xscale{1}\def\yscale{1}}

\def\halfEllipse#1(#2)[#3][#4,#5]{\def\xscale{#4}\def\yscale{#5}\relax
\halfcirc#1(#2)[#3]\def\xscale{1}\def\yscale{1}}

\def\halfdashEllipse(#1)[#2][#3,#4]{\def\xscale{#3}\def\yscale{#4}\relax
\halfdashcirc(#1)[#2]\def\xscale{1}\def\yscale{1}}

\def\halfwaveEllipse(#1)[#2][#3,#4]{\def\xscale{#3}\def\yscale{#4}\relax
\halfwavecirc(#1)[#2]\def\xscale{1}\def\yscale{1}}

\catcode`\@=\the\CatcodeOfAtSign

\title{On the Horava-Lifshitz-like extensions of supersymmetric theories}

\author{M. Gomes}
\affiliation{Instituto de F\'\i sica, Universidade de S\~ao Paulo\\
Caixa Postal 66318, 05315-970, S\~ao Paulo, SP, Brazil}
\email{mgomes,ajsilva@fma.if.usp.br}

\author{J. R. Nascimento}

\affiliation{Departamento de F\'{\i}sica, Universidade Federal da 
Para\'{\i}ba\\
 Caixa Postal 5008, 58051-970, Jo\~ao Pessoa, Para\'{\i}ba, Brazil}
\email{jroberto,petrov@fisica.ufpb.br}

\author{A. Yu. Petrov}

\affiliation{Departamento de F\'{\i}sica, Universidade Federal da 
Para\'{\i}ba\\
 Caixa Postal 5008, 58051-970, Jo\~ao Pessoa, Para\'{\i}ba, Brazil}
\email{jroberto,petrov@fisica.ufpb.br}

\author{A. J. da Silva}
\affiliation{Instituto de F\'\i sica, Universidade de S\~ao Paulo\\
Caixa Postal 66318, 05315-970, S\~ao Paulo, SP, Brazil}
\email{mgomes,ajsilva@fma.if.usp.br}

\begin{abstract}
Within the superfield approach, we formulate two different extensions of the Wess-Zumino model and super-QED with Horava-Lifshitz-like additive terms, discuss their quantum properties and calculate lower contributions to the effective action. In the case of the gauge theory, the one-loop effective potential turns out to be gauge independent.
\end{abstract}

\maketitle

\section{Introduction}

The possibility of Lorentz symmetry breaking is being intensively discussed presently. Following many studies, see e.g. \cite{KostColl}, the Lorentz symmetry breaking can be treated as a natural ingredient of extensions of the standard model. At the same time, supersymmetry is known to be a crucial symmetry of the quantum world due to many remarkable properties of the supersymmetric field theories. In fact, it provides an improvement of the renormalization behaviour in supersymmetric theories due to the well-known mutual cancellation of bosonic and fermionic contributions. Furthermore, the superfield formalism allows a very compact treatment of supersymmetric field theory models (for a general review on supersymmetry see e.g. \cite{BK0,SGRS}). Therefore, a natural problem consists in formulating Lorentz-breaking supersymmetric field theories.

There are essentially two ways to do it up to now. The first is the deformation of the supersymmetry algebra through a Lorentz symmetry breaking implemented by a constant tensor into the anticommutation relations of the supersymmetry generators (known as the Kostelecky-Berger approach \cite{BK}). This approach has been applied to the study of supersymmetric field theories in \cite{CM}, where such models have been studied at the tree level, and in \cite{ourSUSYLV}, where some studies of the the perturbative aspects of these theories have been carried out. This approach is rather universal and can be applied to a wide class of theories, allowing to obtain their CPT-even Lorentz-breaking extensions. The second way to construct Lorentz-breaking theories consists in the introduction of an extra superfield, with some of its components depending on a constant vector (or tensor), as it has been done in \cite{Hel}. We note that in this second method a supersymmetric extension of the Carroll-Field-Jackiw term has been successfully constructed.

None of these methods allow, up to now, to study the supersymmetric extension of the Horava-Lifshitz-like (HL-like) theories characterized by a strong asymmetry between time and space coordinates \cite{Hor}. The reason for that is the presence of different orders in time and space derivatives, in the classical HL-like theories; such a difference apparently cannot be introduced within the two already mentioned approaches. As it was already noted by some of us in an earlier paper \cite{ourSUSYLV}, a natural way of introducing a Lorentz breaking in a superfield model consists in inserting "by hand" extra spatial derivatives in the classical action. Some studies in this direction were performed in \cite{Redigolo}.  Following this line, in this paper, we add to the Wess-Zumino and gauge theory actions extra terms involving higher spatial derivatives of the superfields.

The structure of the paper is as follows. In the section 2, we consider the HL-like extensions for the Wess-Zumino model, with higher derivatives added either to the general or to the chiral Lagrangians and calculate the one-loop low-energy effective actions. In the section 3, we formulate the HL-like extension of the supersymmetric QED, and in the summary we discuss our results.   

\section{The HL-like extensions of the Wess-Zumino model}

Let us treat the simplest supersymmetric theory, that is, the Wess-Zumino model. To do it, let us first define the supersymmetry transformations and generators. We follow here the definitions and conventions of \cite{BK0}.
As usual, the supersymmetry transformations are $\delta\Phi=i(\epsilon^{\alpha}Q_{\alpha}+\bar{\epsilon}_{\dot{\alpha}}\bar{Q}^{\dot{\alpha}})\Phi$, with $Q_{\alpha}$, $\bar{Q}_{\dot{\alpha}}$ being the supersymmetry generators satisfying the relations
\bea
\label{gencom}
\{Q_{\a},\bar{Q}_{\ad} \}&=&2i\sigma_{\a\ad}^m\partial_m ;\quad\,
\{Q_{\a},Q_{\b} \}=\{\bar{Q}_{\ad},\bar{Q}_{\bd}\}=0;\quad\,
[ Q_{\a},\partial_{m} ]=0.
\eea
In the most used symmetric representation they look like 
\bea
\label{gen}
\bar{Q}_{\ad}=i(\frac{\pa}{\pa\bar{\q}^{\ad}}-
i\q^{\a}(\sigma^m)_{\a\ad}\pa_m), \quad\,
Q_{\a}=i(\frac{\pa}{\pa\q^{\a}}+i\bar{\q}^{\bd}(\bar{\sigma}^m)_{\bd\a}\pa_m).
\eea
{ The super covariant derivatives $D_{\alpha}$, $\bar{D}_{\ad}$ are defined to anticommute with these generators, that is, $\{D_{\alpha},Q_{\beta}\}=0$, etc.:
\bea
\label{der}
\bar{D}_{\ad}=(\frac{\pa}{\pa\bar{\q}^{\ad}}+
i\q^{\a}(\sigma^m)_{\a\ad}\pa_m), \quad\,
D_{\a}=(\frac{\pa}{\pa\q^{\a}}-i\bar{\q}^{\bd}(\bar{\sigma}^m)_{\bd\a}\pa_m).
\eea
The chiral superfield $\Phi$ is defined to satisfy the definition $\bar{D}_{\ad}\Phi=0$, and $\bar{\Phi}$ is an antichiral one. Its dynamics is described by the Wess-Zumino model whose action in terms of the} superfields is
\bea
S=\int d^8z\Phi\bar{\Phi}+\int d^6z (\frac{m}{2}\Phi^2+\frac{\lambda}{3!}\Phi^3)+h.c.
\eea
Here, $d^6z=d^4xd^2\theta$ is a chiral superspace measure, and $\int d^8z=d^4xd^4\theta=d^4xd^2\theta d^2\bar{\theta}$ is a whole superspace measure. These definitions will be used thorough this paper.
Let us try to implement a Horava-Lifshitz (HL)-like extension of this theory. 

\subsection{HL-like extension of the general Lagrangian}

Since the term involving the integral over chiral (antichiral) subspace projected into components does not involve derivatives of any fields, let us, as a first attempt, modify only the 
general (K\"{a}hlerian) Lagrangian. To do it, we add a new term with higher derivatives (but only spatial, to avoid the arising of ghost states). So, the new action compatible with the supersymmetry transformations  is 
\bea
\label{sfigeral}
S=\int d^8z\Phi(1+\rho\Delta^{z-1})\bar{\Phi}+(\int d^6z W(\Phi) +h.c.).
\eea
Here $\rho$ is  some constant with a negative dimension, and $z\geq 2$ is a critical exponent. In the most studied case, $W(\Phi)=\frac{m}{2}\Phi^2+\frac{\lambda}{3!}\Phi^3$.

To get some insights, let us write this action in components defined through the projections
\bea
\label{projchir}
\phi(x)&=&\Phi(z)|;\nonumber\\
\psi_{\alpha}(x)&=&D_{\alpha}\Phi(z)|;\nonumber\\
F(x)&=&\frac{1}{4}D^2\Phi(z))|.
\eea
The superpotential term does not differ from the usual, and we have the following component action:
\bea
\label{compwz}
S&=&\int d^4x \Big[\bar{\phi}\, \Box (1+\rho\Delta^{z-1}) \phi-i\bar{\psi}^{\dot{\alpha}}\sigma^m_{\dot{\alpha}\alpha}\partial_m(1+\rho\Delta^{z-1})\psi^{\alpha}+\bar{F}(1+\rho\Delta^{z-1})F-
\nonumber\\&-&
\frac{m}{2}(\phi F+\frac{1}{2}\psi^{\alpha}\psi_{\alpha}+h.c.)-\frac{\lambda}{2}(\phi^2F+\frac{1}{2}\phi\psi^{\alpha}\psi_{\alpha}+h.c.)
\Big].
\eea
Following the general methodology of study of HL theories (and using the arguments of the renormalization group flow \cite{Hor}), we can consider the ``ultraviolet'' limit where we take into account only highest order spatial derivatives. In this limit the action above goes into
\bea
\label{compwz1}
S_{UV}&=&\int d^4x \Big[\bar{\phi} (\pa^2_0-\rho\Delta^z) \phi-i\bar{\psi}^{\dot{\alpha}}(\sigma^0_{\dot{\alpha}\alpha}\partial_0+\rho\sigma^i_{\dot{\alpha}\alpha}\partial_i\Delta^{z-1})\psi^{\alpha}+\rho\bar{F}\Delta^{z-1}F-
\nonumber\\&-&
\frac{m}{2}(\phi F+\frac{1}{2}\psi^{\alpha}\psi_{\alpha}+h.c.)-\frac{\lambda}{2}(\phi^2F+\frac{1}{2}\phi\psi^{\alpha}\psi_{\alpha}+h.c.)
\Big].\label{model1}
\eea
Here,   we omitted the ``crossed'' terms involving both time and space derivatives (like $\bar{\phi}\pa^2_0\Delta^{z-1}\phi$) which do not appear in the usual definition of the HL-like action. We suggest that we can rule out such terms under some conditions. However, even with that simplification, there is no natural way to attribute an uniform, nontrivial, critical exponent to the model. In fact, the first line of (\ref{model1}) indicates the critical exponent
for the scalar field $\phi$ to be $z$ and $2z-1$ for the spinor $\psi$ which are compatible only if $z=1$. We therefore
may conclude that  the uniformity of the critical exponent restricts  (\ref{model1}) to the Lorentz invariant case $z=1$.

The superfield description of the theory (\ref{sfigeral}) is rather interesting. The propagators are
\bea
&&<\Phi(z_1)\bar{\Phi}(z_2)>=\frac{1+\rho\Delta^{z-1}}{\Box(1+\rho\Delta^{z-1})^2-m^2}\delta^8(z_1-z_2);
\nonumber\\
&&<\Phi(z_1)\Phi(z_2)>=<\bar{\Phi}(z_1)\bar{\Phi}(z_2)>^*=-
\frac{m}{\Box(1+\rho\Delta^{z-1})^2-m^2}(-\frac{D^2}{4\Box})\delta^8(z_1-z_2),
\eea
with $D^2$, $\bar{D}^2$ factors associated with the vertices by the same rules as in the usual Wess-Zumino model.

We can calculate the superficial degree of divergence and the effective potential for the theory (\ref{sfigeral}). In this case, 
for a generic amplitude we compute the ultraviolet degree by letting all loop momenta components go uniformily to infinite. Thus the effective degree of the spinor derivatives are $1/2$ whereas for the propagators we have
  $-2z$ for $<\Phi\bar{\Phi}>$, and $1-4z$ for $<\Phi\Phi>$, $<\bar{\Phi}\bar{\Phi}>$  which for $z=1$ reproduces the usual case. Besides that
one has a factor $2$ for any vertex (due to $D^2$, $\bar{D}^2$ factors), but a factor 1 if the vertex has an external line attached, totaling $2V-E$ with $V$ and $E$ the numbers of vertices and external lines, respectively. All loop integrations contribute $4L$, but this number is reduced by $2L$ due to the contraction of any loop into a point, by the rule $\delta_{12}\bar{D}^2D^2\delta_{12}=16\delta_{12}$.  By putting all this together, we have
\bea
\omega=2-E-2(z-1)P-(2z-1)P_c,
\eea
where $P$ is the number of all propagators, and $P_c$ -- of only $<\Phi\Phi>$, $<\bar{\Phi}\bar{\Phi}>$ ones. This $\omega$ reproduces the usual result for $z=1$ where we have only a wave function renormalization (see e.g. \cite{BK0}). If $z\geq 2$, the theory is finite. 

The one-loop effective potential can be obtained through the following standard summation of supergraphs, cf. \cite{YMEP} ( for a review on the methods of calculating the superfield effective potential, see \cite{BKY,YMEP,ourHD}):

\begin{figure}[htbp] 
\includegraphics[scale=0.8]{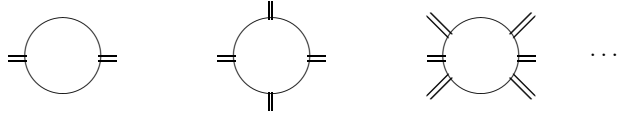}
\caption{\small{Contributions to the one-loop effective potential.}}
\label{figure1}
\end{figure}


The external lines are the alternating $\Psi=-W^{\prime\prime}=m+\lambda\Phi$ (accompanied by $-\frac{\bar{D}^2}{4}$) and the corresponding $\bar{\Psi}$ (accompanied by $-\frac{D^2}{4}$). So, we have the modified Feynman rules:
\bea
&&<\Phi(z_1)\bar{\Phi}(z_2)>=\frac{1}{\Box(1+\rho\Delta^{z-1})}\delta^8(z_1-z_2);
\nonumber\\
&&<\Phi(z_1)\Phi(z_2)>=<\bar{\Phi}(z_1)\bar{\Phi}(z_2)>^*=0.
\eea

For the aims of this paper, we restrict ourselves to the case of the one-loop k\"{a}hlerian effective potential $K^{(1)}$, that is, the effective potential evaluated at the condition that all derivatives of the background super fields are put to zero.
So, similarly to \cite{YMEP}, we have the following sum:
\bea
K^{(1)}=\int d^8z\sum\limits_{n=1}^{\infty}
\frac{1}{2n}\left[
\Psi\bar{\Psi}\frac{\bar{D}^2D^2}{16\Box^2(1+\rho\Delta^{z-1})^2}
\right]^n
\delta^8(z-z^{\prime})|_{z=z^{\prime}}.
\eea
Here the $\frac{1}{2n}$ is a symmetry factor of the corresponding Feynman diagram. We can use the property of the projecting operator: $(\frac{\bar{D}^2D^2}{16\Box})^n=\frac{\bar{D}^2D^2}{16\Box}$ together with the rule for contracting of a loop into a point: $\frac{\bar{D}^2D^2}{16}\delta^4(\theta-\theta^{\prime})|_{\theta=\theta^{\prime}}=1$. So, after the Fourier transform we have
\bea
K^{(1)}=-\int d^4\theta\frac{d^4k}{(2\pi)^4}\frac{1}{k^2}\sum\limits_{n=1}^{\infty}\frac{1}{2n}\left[-\Psi\bar{\Psi}\frac{1}{k^2(1+\rho(-\vec{k}^2)^{z-1})^2}
\right]^n.
\eea
We use the sum: 
\bea
\label{log}
\sum\limits_{n=1}^{\infty}\frac{a^n}{n}=-\ln(1-a),
\eea
so, after the Wick rotation and replacement $\rho\to \rho(-1)^{z-1}$ we have
\bea
\label{k1min}
K^{(1)}=-\frac{i}{2}\int d^4\theta\frac{d^4k}{(2\pi)^4}\frac{1}{k^2}\ln\left[1+\frac{\Psi\bar{\Psi}}{k^2(1+\rho(\vec{k}^2)^{z-1})^2}
\right].
\eea
It is finite for $z>1$ and logarithmically diverges at $z=1$ as it happens in the usual Wess-Zumino model. It is more simple to study $\frac{dK^{(1)}}{d(\Psi\bar{\Psi})}$, to have a purely algebraic expression. Integrating then over $k_0$ (and the corresponding Feynman parameter) by usual procedures of the quantum field theory, we have
\bea
\frac{dK^{(1)}}{d(\Psi\bar{\Psi})}&=&\frac{1}{2}\int\frac{d^3\vec{k}}{(2\pi)^3}\frac{1}{(\vec{k}^2)^{1/2}}
\Big( \vec{k}^2(1+\rho\vec{k}^{2z-2})+\Psi\bar{\Psi}+\nonumber\\ &+&
\sqrt{
(1+\rho
\vec{k}^{2z-2})(\vec{k}^4+\Psi\bar{\Psi}\vec{k}^2+\rho\vec{k}^{2z+2})
}
\Big)^{-1}.
\eea
This expression is evidently finite. Unfortunately, it can be calculated only approximately, by disregarding the subleading powers of $\vec{k}$ and using the approximation $\sqrt{1+x}\simeq 1+x/2$. So, we get
\bea
\frac{dK^{(1)}}{d(\Psi\bar{\Psi})}\simeq \frac{1}{2}\int\frac{d^3\vec{k}}{(2\pi)^3}\frac{1}{(\vec{k}^2)^{1/2}}
\frac{1}{2\rho \vec{k}^{2z}+\frac{3}{2}\Psi\bar{\Psi}}.
\eea
This integral can be found explicitly:
\bea
\frac{dK^{(1)}}{d(\Psi\bar{\Psi})}=\frac{1}{12\pi\Psi\bar{\Psi}} \frac{\csc(\pi/z)}{z}(\frac{4\rho}{3\Psi\bar{\Psi}})^{-1/z}.
\eea
Therefore, the one-loop K\"{a}hlerian effective potential is
\bea
\label{epmin}
K^{(1)}=\frac{1}{12\pi}\csc(\frac{\pi}{z})
(\frac{4\rho}{3})^{-1/z} (\Psi\bar{\Psi})^{1/z}.
\eea
This is an approximate result for the effective potential. It is finite at $z>1$ and diverges at $z=1$ as it must be. Indeed, for $z\to 1$ we can write $z=1+\epsilon$. Since $\csc(\frac{\pi}{1+\epsilon})\simeq \frac{1}{\pi\epsilon}$, so, we have 
\bea
K^{(1)}\simeq \frac{\Psi\bar{\Psi}}{16\pi^2\rho\epsilon}-\frac{\Psi\bar{\Psi}}{16\pi^2\rho}\ln\frac{3\Psi\bar{\Psi}}{4\rho\mu^2},
\eea
 which agrees with the behavior of the usual WZ model \cite{BKY}. The divergence can be cancelled by a simple wave function renormalization. 

\subsection{HL-like extension of the chiral Lagrangian}

An alternative HL-like extension of the Wess-Zumino model would be obtained by adding the HL-like term to the chiral Lagrangian, so, our action will be
\bea
\label{sfields}
S=\int d^8z\Phi\bar{\Phi}+\left[\int d^6z (\frac{1}{2}\Phi(m+a(-\Delta)^z)\Phi+\frac{\lambda}{3!}\Phi^3)+h.c.\right].
\eea

The corresponding superfield propagators are
\bea
<\Phi(z_1)\Phi(z_2)>&=&<\bar{\Phi}(z_1)\bar{\Phi}(z_2)>^*=\frac{m+a(-\Delta)^z}{\Box[\Box-(m+a(-\Delta)^z)^2]}\frac{D^2}{4}\delta^8(z_1-z_2);\nonumber\\
<\Phi(z_1)\bar{\Phi}(z_2)>&=&\frac{1}{\Box-(m+a(-\Delta)^z)^2}\delta^8(z_1-z_2).
\eea
We note that at $a=0$ these propagators reduce to the standard ones of the Wess-Zumino model. There are extra $D^2$, $\bar{D}^2$ factors associated with the vertices just as in the previous case.

We can calculate the one-loop K\"{a}hlerian effective potential.  As usual \cite{YMEP}, the corresponding Feynman supergraphs will represent themselves as rings with alternating chiral and antichiral legs, with any chiral leg accompanied by the factor $-\frac{\bar{D}^2}{4}$, and the antichiral one -- by the factor $-\frac{D^2}{4}$. All propagators are $<\phi\bar{\phi}>$ (with $\phi$, $\bar{\phi}$ being quantum fields), so, we can use again the Fig. 1.

Since the factor $(-\Delta)^z$ acts only in the internal propagators, it is natural to carry out the Fourier transform and include this factor to the definition of the external field; the propagator will be simply $\frac{1}{k^2}$. So, we have the contribution in the form
\bea
\Gamma^{(1)}&=&\sum\limits_{n=1}^{\infty}\frac{1}{2n}\int d^4\theta\int\frac{d^4k}{(2\pi)^4}\left[ (\Psi+a\vec{k}^{2z}) (\bar{\Psi}+a\vec{k}^{2z})
\frac{\bar{D}^2D^2}{16k^4}  \right]^n
\delta(\theta-\theta^{\prime})|_{\theta=\theta^{\prime}},
\eea
where $\Psi=m+\lambda\Phi$.
Then, we use the fact that $[\frac{\bar{D}^2D^2}{16}]^n=(-k^2)^{n-1}\frac{\bar{D}^2D^2}{16}$, contract the loop into a point by the rule $\frac{\{\bar{D}^2,D^2\}}{16}\delta(\theta-\theta^{\prime})|_{\theta=\theta^{\prime}}=1$ and use the identity (\ref{log}).

As a result, after the Wick rotation we obtain
\bea
K^{(1)}=-\frac{1}{2}\int\frac{dk_{0E}d^3\vec{k}}{(2\pi)^4}\frac{1}{k^2}
\ln\Big(
1+\frac{(\Psi+a\vec{k}^{2z})(\bar{\Psi}+a\vec{k}^{2z})}{k^2_{0E}+\vec{k}^2}
\Big).
\eea
The Euclidean momentum square $k^2=k^2_{0E}+\vec{k} ^2$ emerges due to the presence of the $\Box$ because of the usual structure of the spinor supercovariant derivatives.  This integral can be calculated. Indeed, we can rewrite it as
\bea
K^{(1)}=-\frac{1}{2}\int\frac{dk_{0E}d^3\vec{k}}{(2\pi)^4}\frac{1}{k^2_{0E}+\vec{k}^2}
\ln\Big(
k^2_{0E}+\vec{k}^2+(\Psi+a\vec{k}^{2z})(\bar{\Psi}+a\vec{k}^{2z})
\Big).
\eea

We can apply the same approximation as in the previous case, that is, disregard the subleading degrees of momenta, so, we get
\bea
K^{(1)}\simeq- \frac{1}{2}\int\frac{dk_{0E}d^3\vec{k}}{(2\pi)^4}
\frac{1}{k^2_{0E}+\vec{k}^2}
\ln\Big(
k^2_{0E}+\vec{k}^2+\Psi\bar{\Psi}+a^2\vec{k}^{4z}
\Big).
\eea
We can obtain 
\bea
\frac{dK^{(1)}}{d(\Psi\bar{\Psi})}\simeq -\frac{1}{2}\int\frac{dk_{0E}d^3\vec{k}}{(2\pi)^4}\frac{1}{k^2_{0E}+\vec{k}^2}
\frac{1}
{k^2_{0E}+\vec{k}^2+\Psi\bar{\Psi}+a^2\vec{k}^{4z}}.
\eea
We integrate over $k_0$ and $\vec{k}$ as in the previous case. Finally, we get
\bea
K^{(1)}=-\frac{1}{8\pi a^{1/z}}\csc(\frac{\pi}{2z})(\Psi\bar{\Psi})^{1/2z}.
\eea 
We confirm that this effective potential is ultraviolet finite if $z\geq 1/2$, which agrees with the usual Wess-Zumino model. Treating the chiral effective potential, it is easy to find that the first non-trivial contribution to it arises only at the two-loop order just as in the usual Wess-Zumino model \cite{ourchir,West}.
 
\section{The HL-like extension of the supergauge theory}

Now, let us try to introduce the gauge theories within this approach. It is well known (cf. \cite{BK0}) that the free action of the Abelian gauge theory is 
\bea
\label{symact}
S_{SYM}=\frac{1}{64}\int d^6 z\, W^{\a}W_{\a},
\eea
where
\bea
\label{w}
W_{\a}=-\bar{D}^2D_{\alpha}V,
\eea
and the $V(z)$ is a real scalar superfield. 
We can rewrite the action as
\bea
\label{gauge1}
S=-\frac{1}{2}\int d^8z V(-\frac{1}{8} D^{\a}\bar{D}^2 D_{\a})V.
\eea
The action (\ref{symact}) is invariant under gauge transformations
\bea
\delta V=i(\Lambda-\bar{\Lambda}),
\eea
with $\Lambda$ chiral and $\bar{\Lambda}$ antichiral.
It is well known that $(-\frac{1}{8\Box} D^{\a}\bar{D}^2 D_{\a})=\Pi_{1/2}$ is a transverse projector. To perform the Horava-Lifshitz-like extension of this theory in a way compatible with the superfield structure and gauge covariance, we must maintain this projector. So, we deal as in the first example with  the Wess-Zumino model (\ref{sfigeral}), that is, introduce the HL-like operator $1+\rho\Delta^{z-1}$ (so, at $\rho=0$ the usual structure is recovered). Our action takes the form
\bea
S=-\frac{1}{2}\int d^8z V(-\frac{1}{8} D^{\a}\bar{D}^2 D_{\a})(1+\rho\Delta^{z-1})V.
\eea
From the component viewpoint, it looks like
\bea
S=-\frac{1}{4}\int d^4xF_{\mu\nu}(1+\rho\Delta^{z-1})F^{\mu\nu}+\ldots.
\eea
Indeed, this action involves two time derivatives and $2z$ spatial derivatives, together again with some ``crossed'' terms.

We should as usual fix the gauge. It is natural to promote the following extension of the gauge-fixing action:
\bea
S_{GF}=\frac{1}{2\xi}\int d^8 z  V\Pi_0\Box(1+\rho\Delta^{z-1})V, 
\eea
where $\Pi_0=\frac{\{D^2,\bar{D}^2\}}{16\Box}$ is a longitudinal projector (so, $\Pi_{1/2}+\Pi_0=1$, and $\Pi_{1/2}\Pi_0=0$).
As a result, we have the propagator
\bea
<V(z_1)V(z_2)>=-\frac{1}{\Box(1+\rho\Delta^{z-1})}(\Pi_{1/2}+\xi\Pi_0)\delta^8(z_1-
z_2).
\eea
Now, we can implement the coupling of this field with chiral matter. Just as in \cite{highQED}, we can show that there must be only one possible term up to the multiplicative constant factor, that is, 
\bea
S_m=\int d^8z \bar{\Phi}e^{gV}\Phi,
\eea
with no mass or self-coupling terms unless we consider a set of chiral superfields, with the invariant mass matrix $m_{ij}$ and coupling tensor $\lambda_{ijk}$. So, in this case we cannot deform the action of the matter in a HL-like way.

Again as in \cite{highQED}, we can have two types of supergraphs. The first one involves only gauge propagators and quartic vertices in an explicit form:

\vspace*{3mm}

\begin{figure}[htbp] 
\includegraphics[scale=0.8]{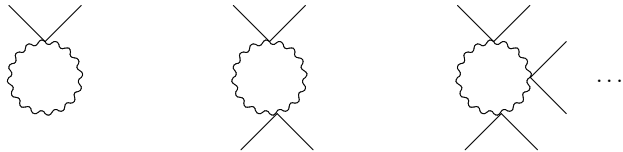}
\caption{\small{Contributions from the supergraphs with quartic vertices only.}}
\label{figure2}
\end{figure}

\vspace*{2mm}

We can perform the sum of these supergraphs:
\bea
K^{(1)}_a=\int
d^8z_1\sum\limits_{n=1}^{\infty}\frac{(-1)^n}{2n}(g^2\Phi\bar{\Phi}
\frac{1}{\Box(1+\rho\Delta^{z-1})}(\Pi_{1/2}+\xi\Pi_0))^n\delta_{12}|_
{\theta_1=\theta_2},
\eea
where $\frac{1}{n}$ is a symmetry factor. 
These diagrams do not involve the triple vertices which will be considered shortly.

By using the properties of the projection operators, we can write
\bea
K^{(1)}_a=\int
d^8z_1\sum\limits_{n=1}^{\infty}\frac{(-1)^n}{2n}(g^2\Phi\bar{\Phi}
\frac{1}{\Box(1+\rho\Delta^{z-1})})^n(\Pi_{1/2}+\xi^n\Pi_0))\delta_{12}|_{\theta_1=
\theta_2}
.
\eea
Since $\frac{D^2\bar{D}^2}{16}\delta_{12}=1$, we have
$\Box\Pi_0\delta_{12}|_{\theta_1=\theta_2}=2$,  and $\Box\Pi_{1/2} 
\delta_{12}|_{\theta_1=\theta_2}=-2$.  Thus, we have
\bea
K^{(1)}_a=\int
d^8z_1\sum\limits_{n=1}^{\infty}\frac{(-1)^n}{n}\frac{1}{\Box}\left[g^2\Phi
\bar{\Phi}\frac{1}{\Box(1+\rho\Delta^{z-1})}\right]^n(1-\alpha^n)\delta^4(x_1-x_2)|_{x_1=x_2
}.
\eea
By carrying out the Fourier transform $\Box\to -k^2$  we arrive at
\bea
K^{(1)}_a=-\int
d^8z\int\frac{d^4k}{(2\pi)^4}\sum\limits_{n=1}^{\infty}\frac{(-1)^n}{n}
\frac{1}{k^2}\left(\frac{g^2\Phi\bar{\Phi}}{k^2(1+\rho(-\vec{k}^2)^{z-1})}\right)^n
(1-\alpha^n).
\eea
Then, by using the expansion (\ref{log}), we have
\bea
\label{k1a}
K^{(1)}_a=\int
d^8z\int\frac{d^4k}{(2\pi)^4}\frac{1}{k^2}\Big[\ln\left(1+\frac{g^2\Phi\bar{\Phi}
}{k^2(1+\rho(-\vec{k}^2)^{z-1})}\right)-\ln\left(1+\frac{\alpha
  g^2\Phi\bar{\Phi}}{k^2(1+\rho(-\vec{k}^2)^{z-1})}\right) \Big].
\eea
Notice that at $\alpha=0$ (Landau gauge), the second term in this
expression  vanishes.

The second type of diagrams involves the triple vertices as well. 
We should first introduce a "dressed" propagator

\vspace*{3mm}

\begin{figure}[ht] 
\includegraphics[scale=0.8]{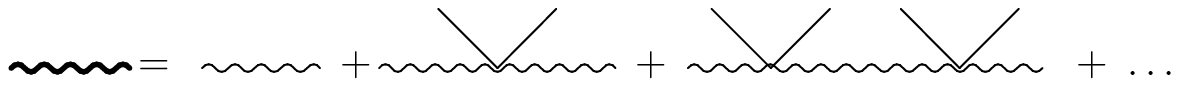}
\caption{\small{''Dressed'' propagator.}}
\label{figure3}
\end{figure}

\vspace*{3mm}

In this propagator, the summation over all quartic vertices is
performed. As a  result, this "dressed" propagator is equal to
\bea
<VV>_D&=&<VV>(1+g^2\Phi\bar{\Phi}<VV>+(g^2\Phi\bar{\Phi}<VV>)^2+\ldots)=
\nonumber\\&=&-
\sum\limits_{n=0}^{\infty}(g^2\Phi\bar{\Phi})^n
\frac{1}{(\Box(1+\rho\Delta^{z-1}))^{n+1}}(\Pi_{1/2}+\alpha\Pi_0)^{n+1}\delta^8(z_1-z_2).
\eea
By summing up, we arrive at
\bea
<VV>_D=-\left(\frac{1}{\Box(1+\rho\Delta^{z-1})+g^2\Phi\bar{\Phi}}\Pi_{1/2}+
\frac{\alpha}{\Box(1+\rho\Delta^{z-1})-\alpha g^2\Phi\bar{\Phi}} \Pi_0\right)\delta^8(z_1-z_2).
\eea

To proceed,  we should sum over diagrams representing themselves as
cycles of  all possible number of links. 
Such diagrams are depicted at Fig. 4.

\vspace*{3mm}

\begin{figure}[ht] 
\includegraphics[scale=0.8]{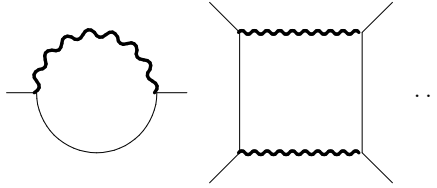}
\caption{\small{Contributions from ''mixed'' supergraphs.}}
\label{figure4}
\end{figure}

\vspace*{3mm}

The complete contribution of all these cycles gives
\bea
K^{(1)}_b=\int
d^8z_1\sum\limits_{n=1}^{\infty}\frac{1}{2n}(g^2\Phi\bar{\Phi}
(<\phi\bar{\phi}>+<\bar{\phi}\phi>)<VV>_D)^n
\delta_{12}|_{\theta_1=\theta_2},
\eea
or, as it is the same,
\bea
K^{(1)}_b=\int
d^8z_1\sum\limits_{n=1}^{\infty}\frac{1}{2n}(g^2\Phi\bar{\Phi}
\Pi_0<VV>_D)^n
\delta_{12}|_{\theta_1=\theta_2}.
\eea
By noting that
\bea
\Pi_0 <VV>_D=-\frac{\alpha}{\Box(1+\rho\Delta^{z-1})-\alpha g^2\Phi\bar{\Phi}}\Pi_0\delta_{12},
\eea
we can rewrite the expression above as
\bea
K^{(1)}_b=\int d^8z_1\sum\limits_{n=1}^{\infty}\frac{(-1)^n}{2n}(
\frac{\alpha g^2\Phi\bar{\Phi}}{\Box(1+\rho\Delta^{z-1})-\alpha g^2\Phi\bar{\Phi}}
)^n \Pi_0
\delta_{12}|_{\theta_1=\theta_2}.
\eea
Since $\Box\Pi_0\delta_{12}|_{\theta_1=\theta_2}=2$, we have
\bea
K^{(1)}_b=\int
d^8z_1\sum\limits_{n=1}^{\infty}\frac{1}{n}\frac{1}{\Box}(-\frac{\alpha
  g^2\Phi\bar{\Phi}}{\Box(1+\rho\Delta^{z-1})-\alpha g^2\Phi\bar{\Phi}})^n\delta^4 
(x_1-x_2)|_{x_1=x_2}.
\eea
By carrying out the Fourier transform and the summation as
above, we 
arrive at
\bea
K^{(1)}_b=\int d^8z\int\frac{d^3\vec{k}dk_{0E}}{(2\pi)^4}\frac{1}{k^2_E}
\ln\Big[1+\frac{\alpha g^2\Phi\bar{\Phi}}{k^2_E(1+\rho(-\vec{k}^2)^{z-1})}\Big].
\eea
By adding this contribution to $K^{(1)}_a$ (\ref{k1a}), we see that
the $\alpha$ dependent contribution vanishes, and the total one-loop
K\"{a}hlerian effective potential is gauge independent, being, after relabeling $\rho(-1)^{z-1}\to \rho$ equal to
\bea
\label{k1a1}
K^{(1)}=\int
d^8z\int\frac{d^3\vec{k}dk_{0E}}{(2\pi)^4}\frac{1}{k^2_E}\ln\Big[1+\frac{g^2\Phi\bar{\Phi}
}{k^2_E(1+\rho(\vec{k}^2)^{z-1})}\Big].
\eea
It is clear that at $\rho=0$, the one-loop K\"{a}hlerian effective potential in the usual supersymmetric gauge theory \cite{YMEP} is re-obtained.

Again, we integrate with use of the same approximations as in the previous cases. We arrive at
\bea
K^{(1)}=\frac{1}{16\pi^2}\csc(\frac{\pi}{z})(\frac{g^2\Phi\bar{\Phi}}{2\rho})^{1/z}.
\eea
This is our final (unfortunately approximate) result. Again, it is finite if $z>1$ and diverges at $z=1$. The same observations made after (\ref{epmin}) can be applied.

\section{Summary}

We have formulated some Horava-Lifshitz-like extensions of the most used superfield theories, that is, the Wess-Zumino model and supersymmetric QED. It turns out that the critical exponent in these theories is not well-defined, however, we can still discuss some of their peculiarities.

We showed that, first, all strength of the superfield approach can be successfully applied to this class of theories, second, the renormalization behaviour of these theories is improved in comparison with their non-supersymmetric analogues, third, we found that, in general, the effective potential may be expressed as integrals over the spatial part of the loop momenta. However, these integrals cannot be expressed in terms of elementary functions. This led us to introduce some approximations to obtain a closed form. 

It is clear that, in principle, this class of theories is not more complicated than the usual Lorentz-invariant supersymmetric field theories. In this respect, we should recall the statement that the theories we discussed flow in the infrared to $z=1$ (or $z=1/2$), that is, to the usual WZ model, and in the ultraviolet to a nontrivial $z$.

{\bf Acknowledgements.} This work was partially supported by Conselho
Nacional de Desenvolvimento Cient\'{\i}fico e Tecnol\'{o}gico (CNPq) and FAPESP. The work by A. Yu. P. has been supported by the
CNPq project No. 303438/2012-6.

\end{document}